\documentclass[twoside]{article}
\usepackage{epsf}
\usepackage{rotating}
\def\1{{\bf 1}}

\def\C{{\bf C}}
\def\R{{\bf R}}
\def\E{{\cal E}}
\def\TF{{\rm TF}}
\def\MTF{{\rm MTF}}
\def\mfr#1/#2{\hbox{${{#1} \over {#2}}$}}
\def\const.{{\rm const.}}
\def\beq{\begin{equation}}
\def\eeq{\end{equation}}
\def\beqa{\begin{eqnarray}}
\def\eeqa{\end{eqnarray}}
\newtheorem{theorem}{Theorem}[section]
\newtheorem{lemma}{Lemma}[section]
\begin{document}
\setlength{\unitlength}{1.0cm}

\title{\bf Quantum dots\\
{\Large \bf A survey of rigorous results}
\footnotetext{\noindent 
{To appear in the proceeedings of the conference on 
{\it Mathematical Results in Quantum Mechanics}, Prague, June 1998}}} 
\author{Jakob Yngvason\\Institut f\"ur Theoretische 
Physik, Universit\"at Wien\\ Boltzmanngasse 5, A 1090 Vienna, 
Austria\\}
\date{}
\maketitle
\section{Introduction}
Modern semiconductor technology has in recent years made it
possible to fabricate ultrasmall structures that confine electrons on 
scales comparable to their de Broglie wavelength.  If the confinement 
is only in one spatial direction such systems are called {\em quantum 
wells}.  In {\em quantum wires} the electrons can move freely in one 
dimension but are restricted in the other two.  Structures that 
restrict the motion of the electrons in all directions are called {\em 
quantum dots}.  The number of electrons, $N$, in a quantum dot can 
range from zero to several thousand.  The confinement length scales 
$R_{1}$, $R_{2}$, $R_{3}$ can be different in the three spatial 
dimensions, but typically $R_{3}\ll R_{1}\approx R_{2}\approx$ 
100 nm.  In models of such dots $R_{3}$ is often taken to be 
strictly zero and the confinement in the other two dimensions is 
described by a potential $V$ with $V(x)\to\infty$ for $|x|\to\infty$, 
$x=(x^1,x^2)\in\R^2$.  A parabolic potential, $V=\mfr1/2\omega |x|^2$, 
is often used as a realistic and at the same time computationally 
convenient approximation.

Quantum dots have potential applications in microelectronics and have 
been extensively studied both experimentally and theoretically.  Apart 
from possible practical uses they are of great interest for basic 
quantum physics.  Their parameters (strength and shape of the 
confining potential, magnetic field strength, number of electrons) can 
be varied in a controlled way and their properties can be studied by 
clever experimental techniques.  This offers many possibilities to 
confront theoretical predictions with experimental findings.  There 
exist by now many excellent reviews on the physics of quantum dots, 
e.g.\ \cite{Chak92}--\cite{Jac98}.  In the 
present contribution the focus will be on some theoretical aspects 
that are only partly covered by these reviews, in particular 
on  rigorous limit theorems \cite{LSY95}, \cite{LSYBi94}
which apply to dots in high magnetic fields and/or with high electron 
density.

A quantum  dot with $N$ electrons is usually modeled by a Hamiltonian of 
the following form, acting on the Hilbert space 
\beq {\cal H}_{N}=\bigwedge\limits^N_1
L^2(\R^2; \C^2)\label{space}\eeq appropriate for two dimensional Fermions of 
spin $1/2$:
\beq H_N = \sum \limits^N_{j=1} H^{(j)}_1 +  \sum
\limits_{1 \leq i < j \leq N} W( x_i - x_j), \label{HN}\eeq
where
$x_i \in \R^2$, $i=1,\dots,N$ and $H^{(j)}_1=1\otimes\cdots 
\otimes H_1\otimes \cdots\otimes 1$ ($H_{1}$ in the $j$-th place) 
with the one-body hamiltonian
\beq H_1 = {\hbar^2 \over 2m_*} \left( {\rm i} \nabla - {e \over \hbar c} 
A(x) 
\right)^2 + V (x)+ g_*
\left( {\hbar e \over 2m_{{\rm e}}c} \right) S_{3}B
-C B.\label{H1}
\eeq
Here $A(x) =\mfr1/2(-Bx^2, Bx^1)$ is the 
vector potential of a homogeneous magnetic field of strength $B$ in 
the $x^3$-direction, $V$ is the confining potential, assumed to be 
continuous with  $V(x)\to\infty$ for $|x|\to\infty$, and $S_{3}$ is the 
spin operator in $x^3$-direction. The parameters 
$m_*$ and $g_*$ are respectively the effective mass and the 
effective $g$-factor of the electrons, while $m_{{\rm e}}$ and $e$ are the 
bare 
values of the electron mass and  electric charge, and $\hbar$ and $c$ have 
their usual meanings. The constant 
\beq C=\left( {\hbar e \over 2m_{{\rm e}} c}
\right) \left({m_{{\rm e}}\over m_*} - {\vert g_*\vert \over 2} \right) \eeq
has been introduced in (3) for convenience: Subtraction of $CB$  
has the effect that the spectrum of the kinetic
energy operator (including spin) $H^{\rm kin}_{1}=H_1-V$ starts at zero for 
all $B$, even 
if $m_{*}\neq m_{{\rm e}}$ and $g_{*}\neq 2$.

The interaction potential $W$ represents the Coulomb repulsion between 
the electrons, modified by the properties of the surrounding medium. 
Usually it is simply taken to be 
\beq W(x_{i}-x_{j})=e_*^2\vert x_{i}-x_{j}\vert^{-1}\label{coul}\eeq
where $e_*=e/\sqrt{\epsilon}$ with $\epsilon$ the dielectric constant, but 
some  regularization of the bare Coulomb 
potential, e.g.,
\beq W(x_{i}-x_{j})=e_*^2\left[(|x_{i}-x_{j}|^2+\delta_{+}^2)^{-1/2}
-(|x_{i}-x_{j}|^2+\delta_{-}^2)^{-1/2}\right]
\label{modcoul}\eeq
with $\delta_{-}>\delta_{+}>0$ \cite{EFK92}, or even a potential that 
depends not just on 
the differences $x_{i}-x_{j}$,  may sometimes fit the effective 
interaction better. For 
the proof of some of the theorems below the important property
of (\ref{coul}) is that $W$ is repulsive, of positive type 
and tends to zero at infinity; these features are shared by 
(\ref{modcoul}). 

Writing the Hamiltonian in the above form is, of course, an 
approximation, because the effect of the medium on the electrons is 
only taken into account through the modification of the parameters 
from their bare values.  The size of a quantum dot ($\approx 100$ nm)
is however usually 
much larger than the lattice constant of the medium where it 
resides ($<0.5$ nm), 
so this approximation is usually a good one.

Quantum dots, especially such with few electrons, are sometimes 
referred to as {\em artificial atoms} with $V$ playing the role of the 
attractive nuclear potential in real atoms.  The analogy is not 
perfect, however, because $V$ is regular around the origin in contrast 
to the potential from an atomic nucleus, and also because the electron 
interaction is the three dimensional Coulomb potential (\ref{coul}) 
(or modified Coulomb potential (\ref{modcoul})), while the motion is 
(essentially) restricted to two dimensions.  But in many respects 
quantum dots can indeed be regarded as artificial atoms, with an 
important additional aspect: The effective parameters are to a certain 
extent tunable and may differ appreciably form their counterparts in 
real atoms.

In a quantum dot the natural atomic unit of length is
$a_*=\epsilon\hbar^2/(m_*e^2)$. Compared with the usual Bohr
radius, $a_0=\hbar^2/(m_{{\rm e}}e^2)=0.53\times 10^{-1}$ nm, the length 
$a_*$ 
is 
typically large, e.g.,
$a_*\approx 185\ a_0\approx 10$ nm in GaAs. The natural energy unit is 
$E_* = e_*^2/a_*=e^4_* m_*/\hbar^2$, and in GaAs $E_*\approx 12$ meV, 
which should be compared with $E_{0}=e^2/a_0=
e^4 m_{{\rm e}}/\hbar^2=27.2$ eV, i.e., $E_*\approx 4\times 10^{-4} E_{0}$.

The natural unit, $B_*$, for magnetic field strength is the 
field at which the magnetic length $\ell_B=\hbar e/(B^{1/2}c)$ equals 
$a_*$, or equivalently, at which the cyclotron energy $\hbar 
eB_{*}/m_{*}c$  equals $E_{*}$.  Hence  $B_*=(a_0/a_*)^2B_0$, where 
$B_0=e^3m_{{\rm e}}^2c/\hbar^3=2.35\times 10^5$ T is the value 
corresponding 
to 
free electrons.  If $a_0/a_*$ is small, $B_*$ can be much smaller than 
$B_0$. Thus $B_*\approx 7$ T in GaAs. This opens the very interesting 
possibility to study in the laboratory magnetic effects, whose analog 
for real atoms require field strengths prevailing only on neutron 
stars.

On the experimental side the main techniques for studying quantum dots 
are {\em charge transport and capacitance spectroscopy} 
(\cite{Be91}, \cite{As96}, \cite{Kouw97}) 
and {\em 
optical far infrared spectroscopy} (\cite{Me93}, \cite{Heit97}).  The former 
is in 
particular suited for measuring the $N$ dependence of ground state 
energies, but excited states can be investigated as well.  The 
applicability of optical spectroscopy is to a certain extent limited 
by Kohn's theorem, to be discussed below, but refined techniques allow 
also to infer many properties by this method. Altogether it is fair 
to say that the energy spectrum of quantum dots and even some aspects 
of the corresponding wave functions are experimentally accessible with 
considerable precision.  See also 
\cite{Zhin97} for  recently discovered effects 
in charge transport spectroscopy that wait for an adequate 
explanation.

On the theoretical side the spectral properties of the Hamiltonian 
(\ref{HN}) 
have been studied by a variety of methods, which can be roughly divided
into the following categories:
\begin{itemize}
\item Exact analytic solutions 
\item Rigorous limit theorems
\item Numerical diagonalizations
\item Hartree and Hartree-Fock approximations
\item Variational calculations 
\item Density functional methods
\end {itemize}

In the condensed matter literature the last four categories are by 
far the most prominent, but the present  survey is mostly concerned 
with the first two, which lie within the realm of mathematical physics. It 
is 
impossible here to do any justice to the extensive physics literature on the 
theory of quantum dots, but an annotated bibliography of some representative 
references  will be given in the last section.
\section{Exact solutions}

\subsection{The Fock-Darwin spectrum}
From now on units are chosen such that
$\hbar=e_{*}=m_{*}=B_{*}=1$. The Hamiltonian (\ref{H1}) can then be written
\beq H_1 = {1\over 2} \left({\rm i} \nabla -  
A(x) 
\right)^2 + V (x)+ \gamma
S_{3}B
-\mfr1/2(1-|\gamma|) B\label{H11}
\eeq
with $\gamma=g_{*}m_{*}/(2m_{{\rm e}})$. For a confining potential of
 the harmonic oscillator form
\beq V(x)=\mfr1/2 \omega |x|^2, \label{quadr}\eeq
the eigenvalues and eigenfunctions of
\beq H_{1}^{\rm orb}=\mfr1/2 \left( {\rm i} \nabla -A(x) 
\right)^2 + \mfr1/2 \omega^2 |x|^2\label{H12}\eeq
were determined by Fock already in 1928 \cite{F28}, and also by 
Darwin in 1930 \cite{D30}. 
The 
Fock-Darwin spectrum
of (\ref{H12}) consists of the eigenvalues
\beq \varepsilon_{k,l}^{\rm FD}=(2k+|l|+1)\Omega-\mfr1/2 l B\eeq
with
\beq \Omega=\left(\mfr1/4 B^2+\omega^2\right)^{1/2},\eeq
$k=0,\,1,\,2\dots$, $l=0,\,\pm1,\,\pm2,\dots$.
The corresponding eigenfunctions are
\beq
\psi_{k,l}^{\rm 
FD}(x)={\rm(const.)}\exp({\rm i}l\varphi)r^{|l|}\exp(-\Omega r^2/2)
L_k^{|l|}(\Omega r^2)
\eeq
where $x=(r\,\cos\varphi,r\,\sin \varphi)$ and $L_k^{|l|}$ is an 
associated Laguerre 
polynomial. These functions are also eigenfunctions of the angular 
momentum $L_{3}$ in the $x^3$-direction with eigenvalue $l$.
Eigenfunctions with the same value of 
\beq n=k+\mfr1/2(|l|-l)\eeq
are grouped together in a {\em Fock-Darwin level} (FDL). In the limit 
$\omega/B\to 0$ the eigenvalues in a FDL coalesce and a FDL becomes 
identical to a {\em Landau level} (LL) with the eigenvalues 
\beq \varepsilon_{n}^{\rm L}=(n+\mfr1/2)B\eeq
and eigenfunctions
\beq
\psi_{k,l}^{\rm L}(x)={\rm(const.)}\exp({\rm i}l\varphi)r^{|l|}
\exp(-B r^2)
L_k^{|l|}(2Br^2)\label{Landau}
\eeq
The degeneracy of a LL per unit area is $B/(2\pi)$.

If the interaction $W$ between the electrons is ignored  the 
FD spectrum together with the Pauli principle, taking spin into 
account,  completely solves 
the eigenvalue problem for $H_{N}$ in the case of the quadratic 
potential (\ref{quadr}). This approximation even fits some experimental data  
quite well \cite{Chak92}, \cite{Ka93}. When the interaction $W$ is taken 
into 
account this 
picture has to be modified, of course.  For a quadratic confining 
potential, however, the FD spectrum continues to apply to the motion of the 
center of 
mass,  independently of the interaction. This simple, but important 
fact \cite{Kohn61}, \cite{GoCho90}
goes under the heading {\em Kohn's theorem}. The proof is essentially 
contained in the identity
\beq N\sum_{j=1}^{N}x_{j}^2=\left(\sum_{j=1}^{N}x_{j}\right)^2
+\sum_{i<j}(x_{i}-x_{j})^2,\label{decoupl}\eeq
for it implies that $H_{N}$ can be written
\beq H_{N}=H_{N}^{\rm CM}+H_{N}^{\rm spin}+H_{N}^{\rm rel}\eeq
where $H_{N}^{\rm rel}$ operates only on the relative coordinates 
$x_{i}-x_{j}$, while all dependence on the center of 
mass coordinate $X=(x_{1}+\cdots+x_{N})/N$ is contained in
\beq H_{N}^{\rm CM}=\mfr1/{2N} \left( {\rm i} \nabla_X -  NA(X) 
\right)^2 + \mfr N/2 \omega^2 |X|^2.\eeq
The spectrum of $H_{N}^{\rm CM}$ is exactly the same as that of $H_{1}^{\rm 
orb}$, independent of $N$ and $W$. In the dipole approximation the 
radiation field couples only to $X$ and in this approximation optical FIR
spectroscopy thus probes only the FD spectrum.

There is a further instance where the FD eigenfunctions play a 
role even for $W\neq 0$.  Let $\Pi_{0}^{\rm FD}$ denote 
the projector on the subspace of ${\cal H}_{N}$ generated by 
eigenfunctions with FD index $n=0$ and complete polarization, 
i.e., spin magnetic moment in the direction of the field.  Consider 
the operator \beq H_{N}'=\Pi_{0}^{\rm FD}H_{N}\Pi_{0}^{\rm FD}.\eeq 
For large $B$ the spectral properties of this operator can be expected 
to approximate those of $H_{N}$.  If $W=0$ the ground state of this 
operator has the orbital wave function 
\beq \Psi^{\rm mdd}=\psi_{0,0}^{\rm FD}\wedge \psi_{0,1}^{\rm FD} 
\wedge\dots 
\wedge\psi_{0,N-1}^{\rm FD}.\eeq 
This eigenfunction is called the {\em maximum density droplet} 
\cite{MacDo93} 
because the electrons are as ``tightly packed'' as possible around the 
origin.  It is an amusing observation  that $\Psi^{\rm mdd}$ 
remains 
an {\em exact eigenfunction} of $H_{N}'$ also for $W\neq 0$.  The 
proof is very simple: $H_{N}'$ and the angular momentum operator $L_{3}$ 
on ${\cal H}_{N}$ commute and have discrete 
spectrum.  Hence for every eigenvalue of $L_{3}$ there is a 
corresponding eigenfunction that is simultaneously an 
eigenfunction of $H_{N}'$.  But the lowest eigenvalue, $N(N-1)/2$, of 
$L_{3}$ 
in 
the subspace of ${\cal H}_{N}$ generated by the FD functions with $n=0$ is 
nondegenerate, and the corresponding eigenfunction $\Psi^{\rm mdd}$ 
must hence also be an eigenfunction of $H_{N}'$. 
There is numerical (\cite{MacDo93}, \cite{MueKoon96}, \cite{Fer97}) and even 
experimental 
\cite{Ost98} evidence that $\Psi^{\rm mdd}$ is a ground 
state of $H_{N}'$  for some range of values of $B$.

\subsection{Analytic solutions for $N>1$}

For $N=2$ the Hamiltonian for the relative coordinate $x=x_{1}-x_{2}$ 
is 
\beq H^{\rm rel}_{2}=({\rm i}\nabla-\mfr1/2A(x))^2+\mfr1/4\omega^2
|x|^2+W(x)\label{H2rel}\eeq
In the case of a pure Coulomb interaction, $W(x)=1/|x|$, explicit 
formulas for eigenfunctions and eigenvalues of (\ref{H2rel})  have been 
found by Taut \cite{Tau95}. His
approach is based on an ansatz for the wave functions of the form
\beq\psi(x)=\exp({\rm i}m\varphi)\exp(-\rho^2/2)\rho^{|m|}P(\rho)
\label{Tautans}\eeq
where $P$ is a polynomial in $\rho=(\Omega/2)^{1/2}r$, with
$\Omega=\left(\mfr1/4 B^2+\omega^2\right)^{1/2}$, 
$\varphi$ is the angular variable and $m\in{\bf Z}$. Solutions of the 
form (\ref{Tautans}) with $P$ a polynomial do not exist for arbitrary 
values of $B$,  but Taut's method produces at least
eigenfunctions and eigenvalues for a countable infinity of values of 
$\Omega$ which accumulate at 0.
In fact, an ansatz like (\ref{Tautans})  in the eigenvalue equation
$H^{\rm rel}_{2}\psi=E\psi$
with $P$ a 
power series, 
\beq P(\rho)=\sum_{\nu=0}a_{\nu}\rho^\nu,\eeq
leads to 
\beq a_{\nu}=F({|m|},\nu,E')a_{0}\eeq
with a certain recursively computable function $F$ and where $E'$ is 
related to $E$ by
\beq E=\mfr1/4\Omega E'-\mfr1/2 mB.\label{EE'}\eeq
The condition that $a_{\nu}$ vanishes for all $\nu$ larger than some 
positive integer
$n$
is equivalent to the two equations
\beq F({|m|},n,E',\Omega)=0\label{T1}\eeq
and
\beq E'=2({|m|}+n).\label{T2}\eeq
For given $n$ and $m$ this gives one or more acceptable values 
for $\Omega$ 
and corresponding energy values 
\beq E=\mfr1/2(n+|m|)\Omega-\mfr1/2 mB.\eeq
The solutions found in this way 
are not necessarily ground states of (\ref{H2rel}) but the position of $E$ 
in the spectrum can be inferred from the number of nodes of the 
corresponding wave function. 

The solutions of Taut seem so far to be the only known exact solutions 
for $N=2$ and the Coulomb interaction (\ref{coul}). They are 
limited to the special values of $\Omega$ defined by (\ref{T1}) and 
(\ref{T2}). For 
$W$ of the inverse 
square form
\beq W(x)=\alpha\vert x\vert^{-2}\label{invsq}\eeq
on the other hand, the Hamiltonian (\ref{H2rel}) can be exactly 
diagonalized fo all $B$ \cite{QuirJoh93}. In fact, addition of 
(\ref{invsq}) merely modifies the centrifugal term in the 
radial part of the FD Hamiltonian (\ref{H12}) 
 and we obtain as eigenvalues of (\ref{H2rel})
\beq E=[2n+\mu+1]\Omega-\mfr1/2 mB\eeq
with $\mu=(m^2+\alpha)^{1/2}$ not necessarily an integer,
and the eigenfunctions
\beq\psi(x)=\exp({\rm 
i}m\varphi)\exp(-\rho^2/2)\rho^{\mu}L_{n}^\mu(\rho^2).
\label{Tautmod}\eeq

It should be noted that the inverse square form (\ref{invsq}) for the 
effective interaction between the electrons is not 
necessarily less 
realistic than the pure Coulomb repulsion (\ref{coul}). In fact, the 
form (\ref{modcoul}) of the interaction, that is motivated by the 
situation in real dots
\cite{EFK92}, has an inverse square decrease for large separation. At 
small separation, on the other hand, the effective interaction may be 
less singular than (\ref{coul}). It is therefore not entirely
academic to consider also a harmonic  interaction \cite{JohPay91} of the 
form
\beq W(x_{i}-x_{j})=2W_{0}-\mfr1/2\beta|x_{i}-x_{j}|^2\label{harm}\eeq
with positive parameters $W_{0}$ and $\beta$. For this case one can even 
solve the problem for {\em all} $N$ exactly,  using (\ref{decoupl}) to 
decouple the 
oscillators. The result for the ground state energy $E^{\rm Q}(N,B)$
of $H_{N}$ with $N\geq 2$ and $\gamma=0$ (for simplicity)
is \cite{JohPay91}
\beq E^{\rm 
Q}(N,B)=\Omega+\mfr1/2(N-1)(N-2)\Omega_{0}(N)-
\mfr1/4N(N+1)B+N(N-1)W_0\eeq
with $\Omega_{0}(N)=(\Omega^2-N\beta^2)^{1/2}$. It is assumed that 
$\omega\geq N^{1/2}\beta$, so $\Omega_{0}(N)\in\R_{+}$ for all $B$.
The corresponding wave function 
for the relative motion is
\beq\psi(x)=\prod_{i<j}[z_{ij}\exp(-\Omega_{0}|z_{ij}|^2/2N)]\eeq
where $z_{ij}=(x_{i}^1-x_{j}^1)-{\rm i}(x_{i}^2-x_{j}^2)$ are the 
relative coordinates regarded as points in $\C$. The form
(\ref{harm}) 
of the repulsion, of course, quite wrong for large separation, but 
this error is to some extent counterbalanced by the confining 
potential.
For a comparison of the solutions for different $W$'s and their 
confrontation with
numerical calculations and experiments
we refer to \cite{Johns95}.

\section{Rigorous limit theorems}

While exact solutions of the eigenvalue problem for $H_{N}$ with $W$ 
of the form (\ref{coul}) or (\ref{modcoul}) are not available for 
$N>2$, it is possible to analyze at least the ground state 
properties for large large $B$ or large $N$ by minimizing simple functionals 
of the electron density.  In this analysis, which implies a drastic 
reduction of the quantum mechanical $N$-body problem, it is not 
necessary that the confining potential $V$ has the quadratic form 
(\ref{quadr}).  
From now on $V$ will stand for an arbitrary continuous function of 
$x\in\R^2$ 
tending to $\infty$ for $|x|\to \infty$; when additional properties 
are required these will be explicitly stated.  It is no restriction to 
assume that $V\geq 0$.  To be able to consider variations of the 
strength of the potential at fixed shape, we write
\beq V(x)=Kv(x)\eeq
with a coupling constant 
$K$.  The interaction $W$ will for definiteness be assumed to be pure 
Coulomb with $e_{*}=1$, i.e.,
\beq W(x-y)=|x-y|^{-1},\eeq
but other repulsive potentials of positive type could be treated 
similarly. The quantum mechanical ground state energy is
\beq E^{\rm Q}(N,B,K)=\langle \Psi_{0},H_{N}\Psi\rangle,\eeq
where $\Psi_{0}$ is a normalized ground state of $H_{N}$. The
corresponding ground state electron density is
\beq\rho^{\rm Q}_{N,B,K}(x)=\sum_{{\rm spins}\,\sigma_{k}=\pm1/2}
\int|\Psi_{0}(x,\sigma_{1};x_{2},\sigma_{2};\dots,
x_{N},\sigma_{N})|^2dx_{2}\cdots dx_{N}.\eeq
We are concerned with  the  asymptotics of these quantities  
when one or more of the parameters $N$, $K$ and $B$ tends  to 
$\infty$ with $v$ fixed.

The large $B$ limit at fixed $N$ and $K$ is easiest and will be 
considered first.

\subsection{High field limit}
In the lowest Landau level (i.e., for $n=0$) 
the wave functions (\ref{Landau}) are 
essentially localized 
on scale $B^{-1/2}$, and the quantum mechanical kinetic energy 
vanishes after the spin contribution and the subtraction of 
$\mfr1/2(1-|\gamma|)$ in (\ref{H11}) have been taken into account. 
In the limit $B\to\infty$ it can therefore  
be expected that a classical model of $N$ point particles with the energy 
function
\beq\E^{\rm P} [x_1,\dots,x_N]= 
\sum_{i=1}^N V(x_i)+\sum_{i<j}|x_{i}-x_{j}|^{-1}\eeq
describes the ground state energy correctly. 
This is indeed the case: Defining
\beq E^{\rm P}(N,K)=\inf \E^{\rm P} [x_1,\dots,x_N]\eeq
we have
\begin{theorem}[High field limit.] For $N$ and $K$ fixed,
\beq \lim_{B\to\infty}E^{\rm Q}(N,B,K)=E^{\rm P}(N,K).\eeq\end{theorem}

{\em Proof:\/} The lower bound for $E^{\rm Q}$ in the proof of this theorem 
is 
trivial: 
Since the kinetic energy part of (\ref{HN}) is $\geq 0$,  $E^{\rm P}\leq 
E^{\rm Q}$. 
The upper bound is a simple variational calculation: Since $V(x)$ tends 
to $\infty$ for $|x|\to\infty$ and $|x_{i}-x_{j}|^{-1}\to\infty$ for 
$|x_{i}-x_{j}|\to 0$,
the infimum  of
$\E^{\rm P}$ is obtained at some point $(\bar x_{1},\dots,\bar 
x_{N})\in\R^{2N}$ with $\bar x_{i}\neq\bar 
x_{j}$ for $i\neq j$.
We test
the Hamiltonian with the wave function
\beq \Psi(x_{1}\dots,x_{N})=\exp({\rm i}x_{1}\times\bar 
x_{1}/B)\psi_{0,0}^{\rm L}(x_{1}-\bar x_{1})\wedge\ldots
\wedge\exp({\rm i}x_{N}\times\bar x_{N}/B)\psi_{0,0}^{\rm L}(x_{N}-\bar 
x_{N})
\eeq
$\wedge$ indicates antisymmetrization in the coordinates 
$(x_{1},\dots,x_{N})$, and the vector product $\times$ of two points 
in $\R^2$ is regarded as $\R$ valued. The spins are totally aligned 
and are therefore omitted in the notation. The kinetic energy of $\Psi$ is 
zero, because 
the wave functions 
$\exp({\rm i}x\times\bar x_{j}/B)\psi_{0,0}^{\rm L}(x-\bar 
x_{j})$ belong to the lowest Landau level. The potential energy part 
approximates $\E^{\rm P} [\bar x_1,\dots,\bar x_N]$ arbitrarily well 
as $B\to \infty$, because 
$|\psi_{0,0}^{\rm L}(x-\bar 
x_{j})|^2$ is essentially localized within a radius $\sim B^{-1/2}$ 
around $\bar x_{j}$ and tends to a delta function as $B\to\infty$.

\subsection{Large $N$ limits}

Before stating the limit theorems formally let us briefly discuss 
their heuristic basis.

For $B=0$ each of 
the $N$ electrons in a dot of radius $R$ occupies a ``private room'' 
of spatial extension $R/N^{1/2}$, because of the Pauli principle.  Hence the 
kinetic energy of an electron is $\varepsilon_{\rm kin}\sim N/R^2$.  
The potential energy due to the confining potential is 
$\varepsilon_{\rm conf}\sim Kv(R)$ and the repulsive energy due to the 
other electrons $\varepsilon_{\rm rep}\sim N/R$.  The radius $R$ can 
be estimated by minimizing the sum $\varepsilon_{\rm kin}+
\varepsilon_{\rm conf}+ \varepsilon_{\rm rep}$. If $K/N$ is kept fixed while 
$N\to \infty$, all these terms 
are proportional to $N$. The radius is therefore {\it independent 
of $N$} an the mean distance between the electrons is $\sim RN^{-1/2}$.  
In this limit the density tends to infinity.  The ground 
state energy of the dot, i.e., the total 
energy $E^{\rm Q}$ of the $N$ electrons, is $\sim N^2$, and this is large 
compared to the exchange/correlation energy $\sim N\cdot 
(N^{-1/2})^{-1}=N^{3/2}$.  One may therefore expect that a 
Thomas-Fermi theory \cite{Shi91} captures the leading asymptotics for 
$N,K\to\infty$ with $K/N$ fixed.

A magnetic field $B$ will not influence the asymptotics as long as 
$B\ll N$, because the energy differences between Landau levels are 
much smaller than the other energy contributions.  If $B\sim N$, on 
the other hand, all Landau levels have to be taken into account, and 
the Thomas-Fermi theory has to be modified (\cite{EFK92},
\cite{EFK93}).  For $B\gg N$ 
the electrons will essentially all sit in the lowest Landau level and 
the kinetic energy contribution vanishes.  The radius is essentially 
independent of $B$ and $N$ (as long as $B\gg N$) and slightly smaller 
than in the $B=0$ case, because the positive kinetic energy term 
$\varepsilon_{\rm kin}\sim N/R^2$ is now missing.  The ground state 
energy is $\sim N^2$ as before.  Theorem 3.1 applies, but the large 
$N$ limit leads to an additional simplification: The point charges can 
be replaced by a continuous distribution \cite{Shi91}, because the 
interelectronic 
distance $\sim RN^{-1/2}$ tends to zero.  The asymptotics is described 
by the energy functional of a charged fluid in the confining potential 
$V$.

For a homogeneous potential $V$ of degree $s\geq 1$, i.e., $V(\lambda 
x)=\lambda^s V(x)$, this last mentioned ``classical'' energy 
functional describes also the large $N$ limit for arbitrary $B$ if 
$K/N\to 0$, in particular if $K$ is fixed while $N\to \infty$.
The radius $R\sim (N/K)^{1/(s+1)}$ 
tends in this case to $\infty$. The density may go to zero or to 
infinity. The kinetic energy per particle $\sim N/R^2\sim N(K/N)^{2/(s+1)}$ 
is in both cases small compared to the other contributions $\sim 
KR^s\sim K(N/K)^{s/(s+1)}$.
Smearing out the point charges brings in an error
$\sim N^{1/2}/R$ per particle, but it is small compared to the total Coulomb 
repulsion $\sim N/R$.

These heuristic considerations will now be turned into precise 
statements.  We define three functionals of the electron density 
$\rho\in L_{1}(\R^2,dx)$, $\rho\geq 0$ as follows.
\smallskip

The {\bf 2D \bf Thomas-Fermi functional} is 
\beq \E^{\TF}[\rho;K]=\pi\int\rho(x)^2 dx
+\int V(x)\rho(x)dx+D(\rho,\rho)\label{TF}\eeq
with $D(\rho,\rho)=\mfr1/2\int\int\rho(x)|x-y|^{-1}\rho(y)dxdy$. All 
integrals are over $\R^2$. The 
$\pi\rho^2$ in the first term is just the kinetic energy density of a 
two dimensional noninteracting electron gas of density $\rho$ at $B=0$.
\smallskip

The {\bf 2D magnetic Thomas-Fermi functional} is defined  as
\beq\E^{\MTF}[\rho;B,K]=\int j_B(\rho(x))dx+\int V(x)\rho(x)dx
+D(\rho,\rho)\label{MTF}\eeq
where $j_B$ is a piece wise linear function representing the kinetic 
energy density of the electrons in a magnetic field $B$, taking all 
Landau levels (including spin) into account. If $\gamma$ in (\ref{H11}) were 
0, the derivative 
$j'_{B}=dj_{B}/d\rho$ would just be the step function
\beq j'_{B}(\rho)=B[2\pi\rho/B]\label{step}\eeq
where $[t]$ denotes the integer part of a real number $t$.
This,
together with $j_{B}(0)=0$ fixes $j_{B}(\rho)$. As $B\to 0$, $j_{B}(\rho)\to 
\pi\rho^2$. If $\gamma\neq 0$, then $j_{B}$ is a slightly more 
complicated step function, cf. Eq. (2.3) in \cite{LSY95}, but since the 
explicit 
form is not significant we refrain form stating it here. When a 
concrete $j_{B}$ is needed for discussion purposes we shall stick to 
the simplest case (\ref{step}).
\smallskip

Finally, the  {\bf classical energy functional} is defined as
\beq\E^{\rm C}[\rho;K]=\int 
V(x)\rho(x)dx+D(\rho,\rho).\label{class}\eeq

All three  functionals are convex, and for each there is a 
unique 
nonnegative density that
minimizes the functional under the constraint $\int\rho(x)dx=N$; this 
is discussed in \cite{LSY95}. 

We denotes the minimizing densities and the corresponding energies by 
$\rho^{\rm TF}_{N,K}$ and $E^{\rm TF}(N,K)$ for the TF functional 
(\ref{TF}), by $\rho^{\rm MTF}_{N,B,K}$ and $E^{\MTF}(N,B,K)$ for the 
magnetic TF functional (\ref{MTF}), and by $\rho^{\rm C}_{N,K}$ and 
$E^{\rm C}(N,K)$ for the classical functional (\ref{class}).  In the 
last case some regularity of $V$ is needed in order to ensure that 
$\rho^{\rm C}_{N,K}$ is a function and not just a positive measure; a 
sufficient conditions is that $V$ satisfies locally an estimate of the 
form 
\beq |\nabla V(x)-\nabla V(y)|\leq {\rm(const.)}\, 
|x-y|^\alpha\eeq with some $\alpha>0 $.

The TF functional (\ref{TF}) and the classical functional 
(\ref{class}) are both limiting 
cases of the MTF functional (\ref{MTF}), for $B\to 0$ and $\to \infty$ 
respectively. More precisely, 
\beqa\lim_{B\to 0} E^{\MTF}(N,B,K)&=&E^{\rm TF}(N,K)\label{first}\\
\lim_{B\to 0}\rho^{\rm MTF}_{N,B,K}&=&\rho^{\rm TF}_{N,K}\eeqa
and 
\beqa\lim_{B\to \infty} E^{\MTF}(N,B,K)&=&E^{\rm C}(N,K)\\
\lim_{B\to \infty}\rho^{\rm MTF}_{N,B,K}&=&\rho^{\rm 
C}_{N,K}.\label{last}\eeqa
The limit for the densities  should be understood in the 
weak $L_{1}$ sense, but for special $V$ much stronger convergence may
hold. For instance, if $V$ is monotonically increasing with $|x|$, 
$\rho^{\rm C}_{N,K}$ is a bounded function, and $\rho^{\rm C}_{N,K}=
\rho^{\rm MTF}_{N,B,K}$ for sufficiently large $B$, because 
$j_{B}(\rho^{\rm C}_{N,K})=0$ for $B>2\pi\Vert \rho^{\rm 
C}\Vert_{\infty}$.

The MTF theory has two nontrivial parameters 
because of the 
{\bf scaling relations}
\beqa E^{\MTF}(N,B,K)&=&N^2E^{\MTF}(1,B/N,K/N)\\
\rho^{\MTF}_{N,B,K}(x)&=&N\rho^{\MTF}_{1,B/N,K/N}(x).\eeqa
Corresponding relations (without $B$) 
hold for the TF theory and the classical theory, and also for $E^{\rm 
P}$.

A further important property of the densities is their compact support:
For fixed $K/N$ the minimizers of  $\E^{\TF}$, $\E^{\MTF}$, $\E^{\rm 
C}$ and also of $\E^{\rm P}$ 
have support in a disc whose radius is uniformly bounded in $N$ 
and $B$ (Lemma A.1 in \cite {LSY95}).

 Each minimizer satisfies a variational equation, which in the case of the 
MTF theory has an unusual form, since it consists really of inequalities. To 
state it compactly it is convenient to modify the definition (\ref{step}) 
slightly and regard $j'_{B}$ as an {\em interval valued} function if 
$2\pi\rho/B$ is an integer, namely, if $2\pi\rho/B=n$, then $j'_{B}$ 
is the closed interval
$[(n-1)B,nB]$. The {\bf MTF equation} that is satisfied by $\rho^{\MTF}$ can 
then be written 
\beq \mu-V(x)-\rho*\vert x\vert^{-1} \left\{ 
\begin{array} {r@{\quad \hbox{\rm if}\quad} l} 
	\in j'_B(\rho(x))&
\rho(x)>0\\
\leq 0 & \rho(x)=0\end{array}\right.\eeq
with a unique $\mu=\mu(N,B,K)$.
Such generalized variational equations have been studied by Lieb and 
Loss \cite{LL}.

If the potential is quadratic,  $V(x)=K|x|^2$, there is an explicit formula 
(\cite{Shi91}, \cite{LSY95}) for the minimizer for $\E^{\rm C}$, which is 
equal to $\rho^{\rm MTF}$
for $B$ sufficiently large:
\beq \rho^{\rm C}_{N,K}(x) = \left\{ 
\begin{array} {r@{\quad \hbox{\rm if}\quad} l} 
	{3 \over 2 \pi} N\lambda \sqrt{1 -
\lambda \vert x
\vert^2} & \vert x \vert \leq \lambda^{-1} \\
0 & \vert x \vert > \lambda^{-1}\end{array}\right.\label{rhoc}\eeq
with $\lambda = (8K/3\pi N)^{2/3}$.
The density profile has the shape of a half ellipsoid with a maximum 
at $x=0$. Note the difference between the two dimensional case considered 
here, 
and three dimensional electrostatics: In three dimensions the density 
would be homogeneously distributed in a ball.

The criterion for $\rho^{\rm MTF}=\rho^{\rm C}$ is that 
$j_{B}(\rho^{\rm C}(0))=0$, which holds if 
\beq B\geq (6/3^{2/3}\pi^{5/3})K^{2/3}N^{1/3}.\label{cond}\eeq

Numerically computed profiles of the minimizers $\rho^{\MTF}$ and
the corresponding effective potentials 
\beq V_{\rm eff}(x)=V(x)+\rho^{\MTF}*\vert x\vert^{-1}\eeq
with $V(x)=K|x|^2$ are shown in Fig.\ 1. The computations were carried 
out by Kristinn Johnsen. 

At the highest value of the 
field (Fig.\ 1(a)) condition (\ref{cond}) is fulfilled and $\rho^{\rm 
MTF}$ has the form (\ref{rhoc}). On the support of 
$\rho^{\MTF}=\rho^{\rm C}$ we have $V_{\rm eff}(x)=$constant$=\mu$.

\begin{figure}
\begin{center}
\begin{picture}(12.5,12.5)
%
%
\epsfysize=3.0cm
\put(-0.3,0.5){\epsfbox{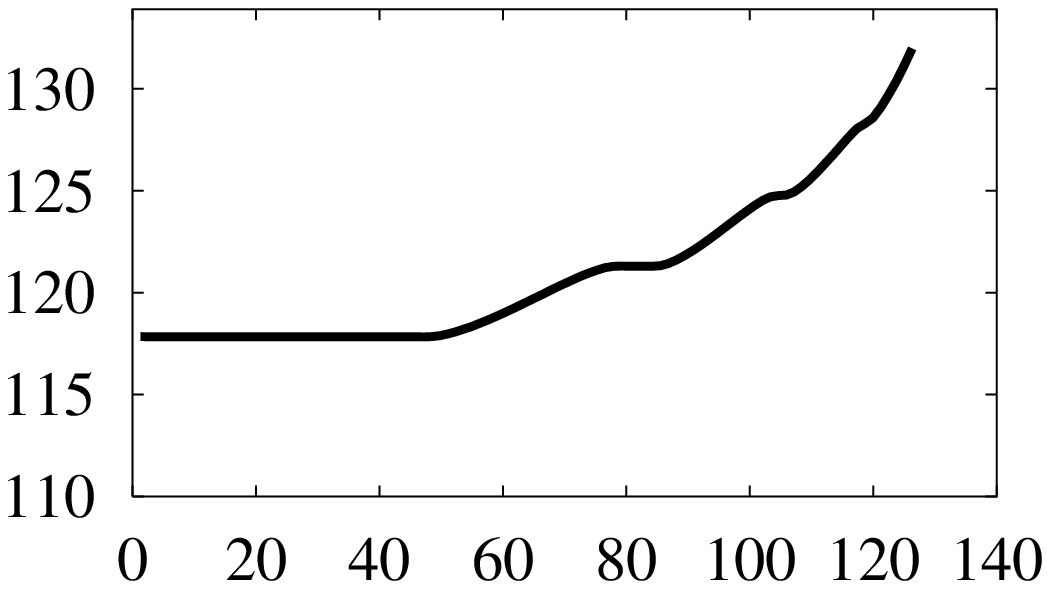}}
\epsfysize=3.0cm
\put(-0.3,3.0){\epsfbox{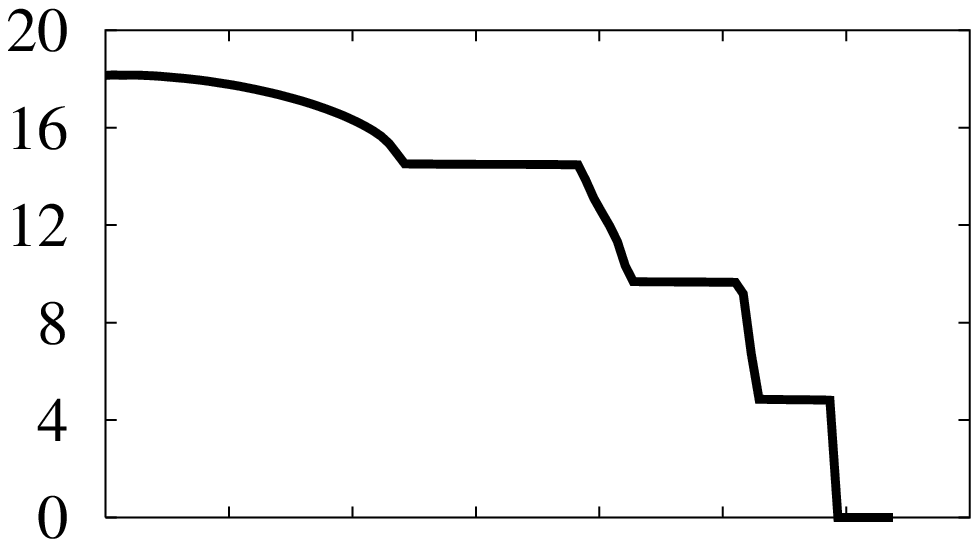}}
\put(0.0,1.3){\rotatebox{90}{$V_{{\rm eff}}$ $({\rm meV})$}}
\put(0.0,3.7){\rotatebox{90}{$\rho$ $(10^{14}\mbox{m}^{-2})$}}
\put(2.7,.1){$r$ (\mbox{nm})}
\put(4.7,5.2){(c)}
\epsfysize=3.0cm
\put(5.95,0.5){\epsfbox{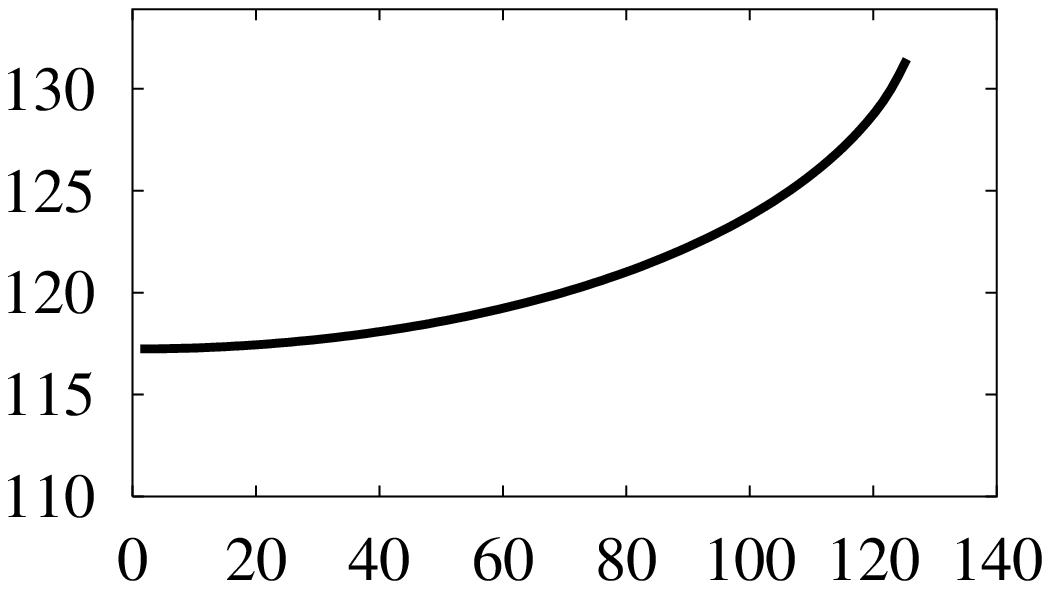}}
\epsfysize=3.0cm
\put(5.95,3.0){\epsfbox{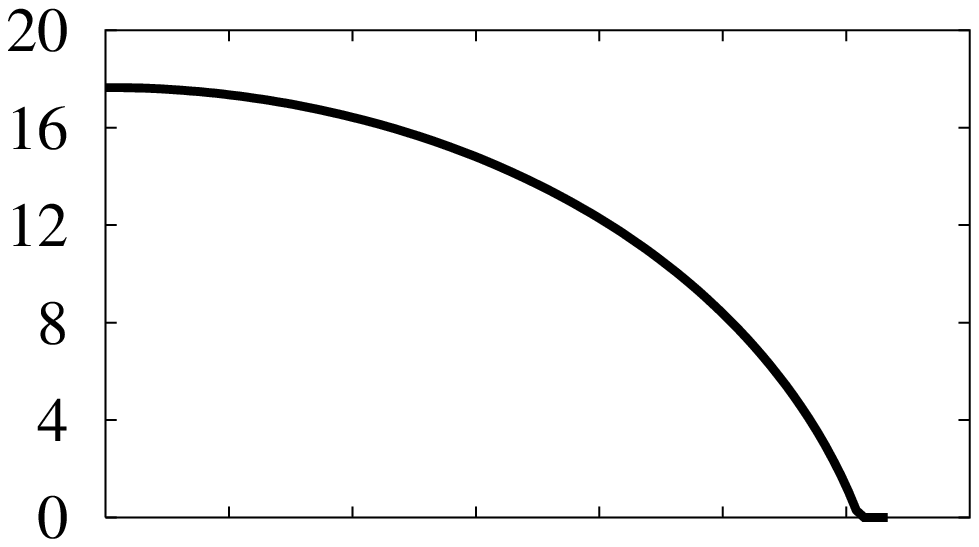}}
\put(6.25,1.3){\rotatebox{90}{$V_{{\rm eff}}$ $({\rm meV})$}}
\put(6.25,3.7){\rotatebox{90}{$\rho$ $(10^{14}\mbox{m}^{-2})$}}
\put(8.95,.1){$r$ (\mbox{nm})}
\put(10.95,5.2){(d)}
\epsfysize=3.0cm
\put(5.95,6.9){\epsfbox{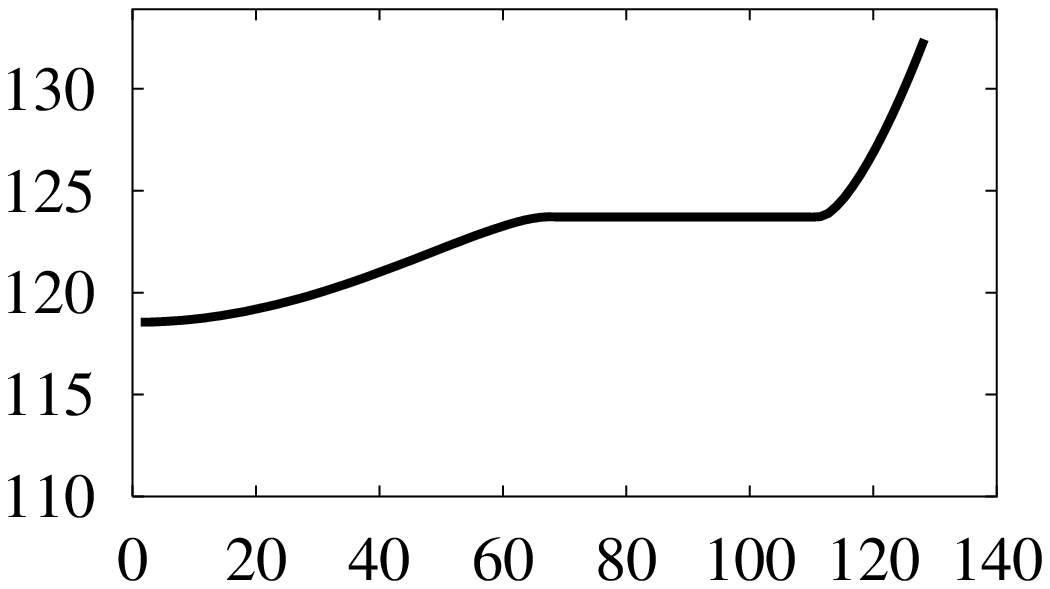}}
\epsfysize=3.0cm
\put(5.95,9.4){\epsfbox{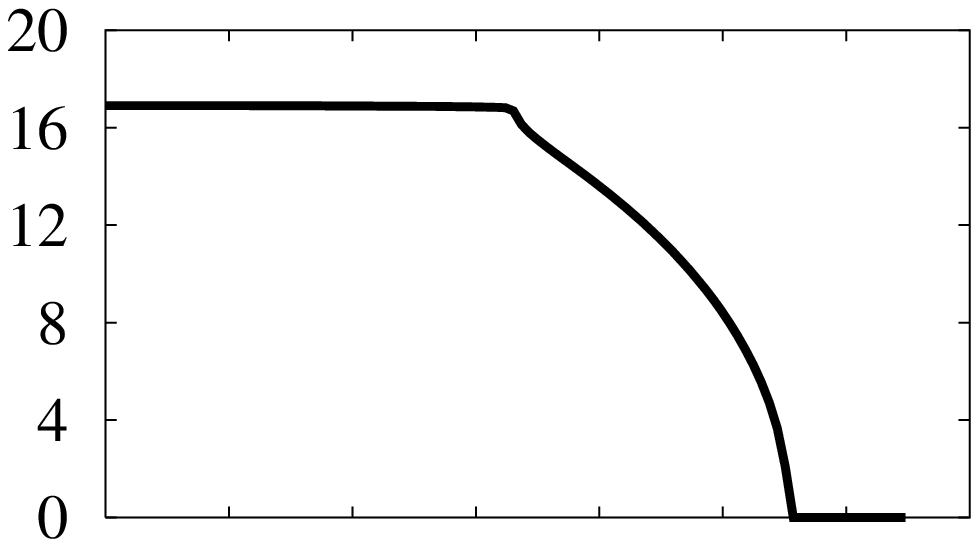}}
\put(6.25,7.7){\rotatebox{90}{$V_{{\rm eff}}$ $({\rm meV})$}}
\put(6.25,10.1){\rotatebox{90}{$\rho$ $(10^{14}\mbox{m}^{-2})$}}
\put(8.95,6.5){$r$ (\mbox{nm})}
\put(10.95,11.6){(b)}
\epsfysize=3.0cm
\put(-0.3,6.9){\epsfbox{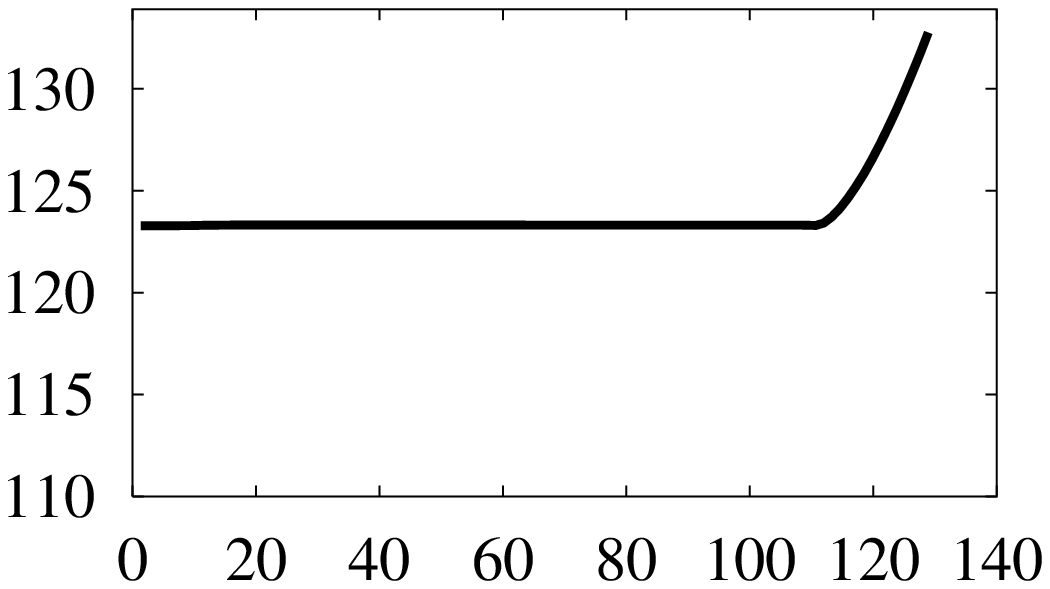}}
\epsfysize=3.0cm
\put(-0.3,9.4){\epsfbox{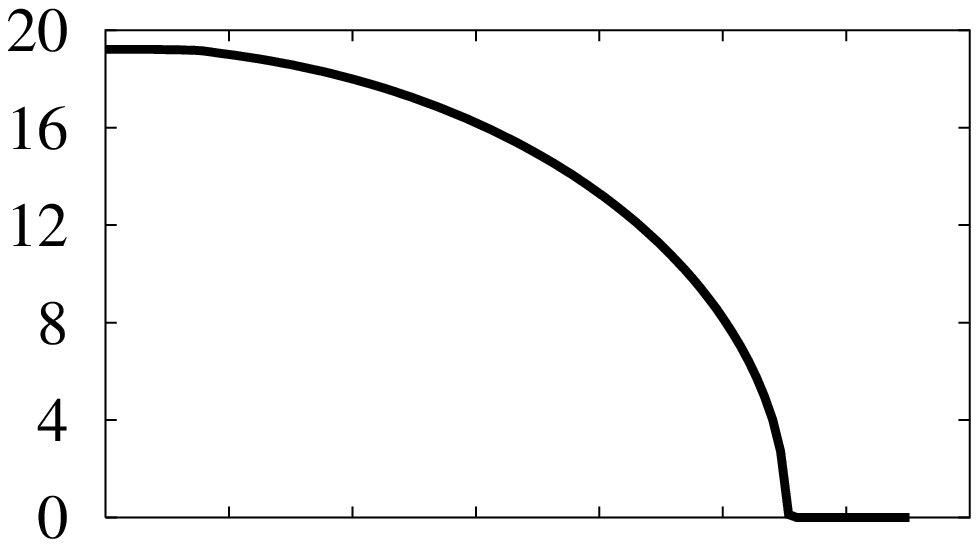}}
\put(0.0,7.7){\rotatebox{90}{$V_{{\rm eff}}$ $({\rm meV})$}}
\put(0.0,10.1){\rotatebox{90}{$\rho$ $(10^{14}\mbox{m}^{-2})$}}
\put(2.7,6.5){$r$ (\mbox{nm})}
\put(4.7,11.6){(a)}
\end{picture}
\end{center}
\caption{
Density profiles and effective potentials for the MTF theory at 
different magnetic field strengths, calculated for $N=50$ and 
$V(x)=K|x|^2$ with $K=1,7$ meV and the material parameters of 
GaAs. (a) $B=8$ T, (b) $B=7$ T, (c) $B=2$ T, (d) $B=0$ T.}
\end{figure}

When the field is gradually turned down the maximal density $B/(2\pi)$ of 
electrons that can be accommodated in the lowest Landau level goes down 
also.  Condition (\ref{cond}) no longer holds, i.e., the density 
$\rho^{\rm C}$ near the center is higher than $B/(2\pi)$ and charges 
have to be moved into other states in phase space.  If $B$ is only 
slightly smaller than the value given by (\ref{cond}) (Fig.\ 1(b)) it would 
cost too much energy to bring the electrons near the origin into the 
next Landau level and it pays to move them spatially away from the 
center, because the potential energy increase is less than $B$.  
Hence in a certain 
range of $B$ values, the density near the center is locked at the 
value $B/(2\pi)$ ( ``incompressible'' domain).  The effective 
potential is no longer constant in this domain.  In the complementary 
``compressible'' domain, on the other hand, the density is below the 
critical value $B/(2\pi)$, and tends to zero in such a way that the 
effective potential stays constant.  Reducing the field 
strength further brings more Landau levels into play (Fig.\ 1(c)).  
Incompressible domains, where the density is an 
integer multiple of $B/(2\pi)$, alternate with compressible domains, 
where the effective potential has a constant value.  When $B\to 0$ the 
profile becomes indistinguishable from the smooth profile of 
$\rho^{\rm TF}$ (Fig.\ 1(d)).
It is interesting to note that the alternation of compressible and 
incompressible domains in moderate magnetic fields 
may account for some fine structure in the charge
transport spectroscopy of quantum dots with a large number of 
electrons \cite{VRK94}.

The basic {\bf limit theorem} \cite{LSY95} that relates the energy 
functionals 
(\ref{TF})-(\ref{class}) to 
the quantum mechanical ground state of $H_{N}$ is as follows:

\begin{theorem}[High density limit.] Let $N\to\infty$ with $K/N$ fixed. 
Then,
uniformly in $B/N$, 
\beq E^Q(N,B,K)/E^{\MTF}(N,B,K)\rightarrow 1\label{elim}\eeq
and 
\beq N^{-1}\rho^Q_{N,B,K}(x)\rightarrow \rho^{\MTF}_{1,B/N,K/N}(x)\eeq
in weak $L^1$ sense.
Moreover, if $B/N\to 0$, then
\beq E^Q(N,B,K)/E^{\TF}(N,B,K)\rightarrow 1\eeq
\beq N^{-1}\rho^Q_{N,B,K}(x)\rightarrow \rho^{\TF}_{1,B/N,K/N}(x),\eeq
and if $B/N\to\infty$, then
\beq E^Q(N,B,K)/E^{\rm C}(N,B,K)\rightarrow 1\eeq
\beq N^{-1}\rho^Q_{N,B,K}(x)\rightarrow \rho^{\rm 
C}_{1,B/N,K/N}(x).\eeq
\end{theorem}

According to this theorem there are thus three asymptotic regimes for 
quantum 
dots as $N$ and $K$ tend to $\infty$ with $K/N$ fixed: $B\ll N$, $B\sim N$ 
and 
$N\ll B$. This should be compared with the more complex situation for three 
dimensional natural atoms in strong magnetic field, where there are five 
regimes
\cite{LSY94a}, \cite{LSY94b} for $N\to\infty$ with $Z/N$ fixed ($Z$ = 
nuclear 
charge): $B\ll N^{4/3}$, $B\sim  N^{4/3}$, $N^{4/3}\ll B\ll N^3$, $B\sim 
N^3$, 
$ N^{3}\ll B$.
\smallskip

For homogeneous potentials a stronger asymptotic result holds, for $K/N$ may 
tend 
to zero as $N\to\infty$.
\begin{theorem} [Homogeneous potentials.]
Assume that $V$ is homogeneous of degree $s\geq1$, i.e.,
\beq 
	V(\lambda x)=\lambda^sV(x).
\eeq
Then
\beq \lim_{N\to\infty} E^{\rm Q}(N,B,K)/E^\MTF(N,B,K)=1\eeq
uniformly in $B$ and in $K$ as long as $K/N$ is bounded above.
Moreover, if $K/N\to 0$ as $N\to\infty$, then
\beq \lim_{N\to\infty} E^{\rm Q}(N,B,K)/E^{\rm C}(N,K)=1\eeq
uniformly in $B$.
\end{theorem}

We shall now discuss briefly the main techniques used for the proof of these 
theorems. As usual it is sufficient to prove the limit theorems for the 
energy, because the corresponding results for the density can be obtained by 
variation with respect to the potential $V$. The basic result is thus Eq.\ 
(\ref{elim}); the other limit theorems follow by 
(\ref{first})--(\ref{last}). 
One has to prove upper and lower bounds for the quantum mechanical energy 
$E^{\rm Q}$ in terms of the energy $E^{\MTF}$, with controllable errors. 

The upper bound is obtained, using the variational principle of \cite{Lvar}, 
by testing $H_N$ with a suitable one particle density operator. Its kernel 
in 
the space and spin variables $x,\sigma$ has the form
\beq {\cal K}(x,\sigma;x',\sigma')=\sum_\nu f_\nu(u)\Pi_{\nu 
u}(x,\sigma;x',\sigma')d^2u\eeq
where the sum is over all Landau levels and $f_\nu(u)$ is the filling factor 
of the $\nu$-th Landau level at point $u$ when the density is 
$\rho^{\MTF}(u)$. The kernel $\Pi_{\nu u}(x,\sigma;x',\sigma')$ is obtained 
from the kernel $\Pi_\nu(x,\sigma;x',\sigma')$ of the projector on the 
$\nu$-th Landau level by localizing around $u$ with a smooth function $g$ of 
compact support, 
i.e.,
\beq \Pi_{\nu 
u}(x,\sigma;x',\sigma')=g(x-u)\Pi_\nu(x,\sigma;x',\sigma')g(x'-u).\eeq
This operator is positive and approximately a projector, localizing 
simultaneously in space, i.e., around $u$, and in the Landau level index 
$\nu$. 
By letting the support of $g$ shrink with $N$ more slowly that the average 
electron spacing $N^{-1/2}$, the error terms in the estimate above for 
$E^{\rm 
Q}-E^{\MTF}$ are of lower order than $N^2$, which is the order of 
$E^{\MTF}$.

The lower bound for $E^{\rm Q}$ is proved separately for large $B$ and for 
small $B$. For large $B$, i.e., $B\gg N$,  one starts with the obvious 
estimate $E^{\rm Q}\geq E^{\rm P}$. One then has to compare $E^{\rm P}$ with 
$E^{\rm C}$, i.e., the energy of point charges with those of smeared 
charges. 
Since the electron distance is $\sim N^{-1/2}$ the self energy of a smeared 
unit charge is $\sim N^{1/2}$. Hence an estimate 
\beq E^{\rm P}(N,K)\geq E^{\rm C}(N,K)-bN^{3/2}\eeq
 with $b$ depending only on $K/N$ is to be expected, and this can indeed be 
proved, using an electrostatic lemma of Lieb and Yau \cite{LY88}.

The lower bound for small $B$,  i.e., $B\ll N$ or $B\sim N$, 
requires an estimate on 
the 
indirect Coulomb energy, that is derived in essentially the same way as a 
corresponding inequality in \cite{L79}, using the positive definiteness of 
the 
Coulomb interaction \ref{coul}, cf. also \cite{Ba92}.
\begin{lemma}[Exchange inequality in 2 dimensions.]
\beq \sum_{\rm spins}\,\, \int \limits_{\R^{2N}} \vert \Psi \vert^2
\sum
\limits_{i < j } \vert x_i - x_j \vert^{-1} \geq D(\rho_\Psi,\rho_\Psi) - 
192 (2\pi)^{1/2}\int
\limits_{\R^2}
\rho_\Psi^{3/2}.\label{exchange} \eeq
\end{lemma}
In order to control negative
term $\sim \int \rho_\Psi^{3/2}$ on the right side of  
(\ref{exchange}) a lower bound on the kinetic energy is needed. This 
in turn  is derived from a two 
dimensional magnetic Lieb-Thirring inequality, which has to be proved in a 
slightly different way from the corresponding inequality in \cite{LSY94b}, 
because there is no kinetic energy associated with a motion in the $x^3$ 
direction. The following inequality is adequate for the present purpose, but 
sharper Lieb-Thirring type inequalities that hold even for inhomogeneous 
fields 
have been derived by Erd\H os and Solovej \cite{ESa}, \cite{ESb}.
\begin{theorem} [Lieb-Thirring inequality in 2 dimensions.]
Let $U$ be 
locally integrable
and let
$e_1(U), e_2(U), \ldots$ denote the negative eigenvalues (if any)
of the Hamiltonian
${1\over 2}({\rm i}\nabla-{\bf A})^2+S_{3}B-U$. 
Define $\vert U \vert_+ (x) = \mfr1/2 [\vert U(x)
\vert + U(x)]$.  For all $0<\lambda<1$
we have the estimate
\beq
\sum_j |e_j(U)|\leq  \lambda^{-1}{B \over 2\pi} \int_{\R^2} 
|U|_+
(x) dx+ \mfr3/4(1-\lambda)^{-2} \int_{\R^2} |U|_+^2 (x) dx.\label{LT}
\eeq\end{theorem}
By a Legendre transformation with respect 
to $U$ it follows from (\ref{LT}) that for all $0<\lambda<1$ 
the kinetic energy $T_{\Psi}$ of a state 
$\Psi$ is bounded 
below by 
$\mfr1/3(1-\lambda)^2\int [\rho_{\Psi}-\lambda^{-1}B/2\pi]_{+}^2$. It 
is then possible, for $B/N$ smaller than a certain critical value 
depending on $v$, to chose an $N$ dependent $\varepsilon>0$ in such a 
way that $\varepsilon\to 0$ as $N\to\infty$, but $\varepsilon T_{\psi}-
\hbox{\rm (const)\, }\int
\rho_\Psi^{3/2}\geq 0$ for all $N$-particle states $\Psi$.

\section{Other approaches}

The rigorous results presented above concern mainly (but not 
exclusively)  the extreme cases
of very few ($N=1$ or $N=2$) or very many ($N\to\infty$) electrons. 
These cases play a similar role as the hydrogen atom and the Thomas-Fermi 
atom do in ordinary atomic physics, i.e., they set a standard 
that can be used as a starting point of various approximation schemes, 
or as a test for such schemes whose connection with the original
Hamiltonian (\ref{HN}) may not be entirely clear. The physics literature on 
quantum dots is by now quite extensive, 
and many approaches have been used for gaining 
insight where rigorous results are not yet available. Here
it is only possible to mention the main methods and give a list of 
some references that are representative for the approaches of condensed 
matter 
physicists to these problems and from it further sources can be 
traced. See also \cite{Jac98} for a more 
comprehensive list.

For small $N$ a direct numerical diagonalization of $H_{N}$ is 
possible and has been carried out, e.g., in 
\cite{Yang93}, \cite{Wagn92}, \cite{Pfann94}, \cite{Meza97}, 
\cite{Eto97}, for various values of $N\leq 8$.  It is, of course, 
necessary to restrict $H_{N}$ to a finite dimensional subspace of the 
full Hilbert space, and the error made in this step is seldom 
estimated rigorously.  By the mini-max principle, however, the 
computed values give at least upper bounds to the true eigenvalues.  
One of the features studied by this method is the orbital angular 
momentum and spin of the ground state, and analogs of Hund's rules 
from atomic physics, as well as oscillations between triplet and 
singlet states for $N=2$ as $B$ is varied have been seen in the 
calculations \cite{Chak92}, \cite{Wagn92}.

For $N>10$ numerical diagonalization of the Hamiltonian is at present 
hardly feasible and resource is taken to other techniques of many body 
theory like Hartree and Hartree-Fock approximations (e.g., 
\cite{Pfann94}, \cite{Pfann95}, \cite{MueKoon96}, \cite{Heit97}), 
perturbation theory \cite{Anis98}, variational methods (e.g. 
\cite{DiNaz97}), quantum Monte Carlo methods (\cite{HarNie98a}, 
\cite{HarNie98b}), and (current) density functional theory 
(\cite{Fer94}, \cite{Fer97}, \cite{Pi98}, \cite{Hein98}).  In quantum 
dots with a moderate electron number correlations play a 
much greater role than in natural atoms because the confining potential 
is usually quite shallow around the origin and the density may be low.  
(By contrast, one of the main steps in the proof of Theorem 3.2 is to 
show that exchange and correlation effects vanish in the high density 
limit for arbitrary magnetic fields.) Strong corelations together 
with strong dependence on magnetic fields make up much of the special 
flavor of quantum dot physics and it remains a challenge for 
mathematical physics to capture these effects in a rigorous way.


\begin{thebibliography}{99}
\bibitem[1]{Chak92} T.\ Chakraborty, {\em Physics of the Artificial Atoms: 
Quantum Dots in a Magnetic Field}, Comments Cond.\ Mat.\ Phys. {\bf 
16}, 35-68 (1992)


\bibitem[2]{Ka93} M.A.\ Kastner, {\em Artificial atoms}, 
Phys.\ Today
{\bf 46}, 24--31 (1993)

\bibitem[3]{Heit93} D.\ Heitmann and J.\ Kotthaus, {\em The spectroscopy of 
quantum dot 
arrays}, 
Phys.\ Today
{\bf 46}, 56--63 (1993)

\bibitem[4]{Johns95} N.\ F.\ Johnson, {\em Quantum dots: few-body, low 
dimensional systems}, J.\ Phys.: Condens.\ Matter {\bf 7}, 965-989 
(1995)

\bibitem[5]{Ka96} M.A.\ Kastner, {\em Mesoscopic Physics with 
Artificial Atoms}, 
Comments Cond.\ Mat.\ Phys. {\bf 
17}, 349--360 (1996)

\bibitem[6]{As96}R.C.\ Ashori, {\em Electrons on artificial atoms}, 
Nature {\bf 379}, 413--419 (1996)
	
\bibitem[7]{JeHa97} J.\ H.\ Jefferson and W.\ H\"ausler, {\em Quantum dots 
and artificial atoms}, Molecular Physics Reports, {\bf 17} 81-103 (1997)

\bibitem[8]{Jac98}L.\ Jacak, P.\ Hawrylak, A.\ W\'ojs, {\em Quantum Dots}, 
Springer, Berlin etc., 1998

\bibitem[9]{LSY95} E.H.\ Lieb, J.P.\ Solovej and J.\ Yngvason, {\em The 
Ground States of Large Quantum Dots in Magnetic Fields}, Phys.\ Rev.\ B 
{\bf 51},
10646--10665 (1995)

\bibitem[10]{LSYBi94} E.H.\ Lieb, J.P.\ Solovej and J.\ Yngvason, {\em 
Quantum Dots}, in: Proceedings of the Conference on Partial 
Differential Equations and Mathematical Physics, University of 
Alabama, Birmingham, 1994, I. Knowles, ed., pp.\  157--172, 
International Press 1995

\bibitem[11]{EFK92} P.L.\ McEuen, E.B.\ Foxman, J.\ Kinaret, U.\ Meirav, 
M.A.\ 
Kastner, N.S.\ Wingreen and S.J.\ Wind, {\em Self consistent addition 
spectrum of a Coulomb island in the quantum Hall regime}, Phys.\ Rev.\ B 
{\bf 45}, 11419--11422 (1992) 

\bibitem[12]{Be91} C.\ W.\ J.\ Beenakker, {\em Theory of Coulomb-blockade 
oscillations in the conductance of a quantum dot}, Phys.\ Rev.\ B {\bf 
44}
1646--1656 
(1991)

\bibitem[13]{Kouw97}L.\ P.\  Kouwenhouven, T.H.\ Osterkamp, M.\ W.\ S.\ 
Danoesastro, M.\ Eto, D.\ G.\ Austing, T.\ Honda and S.\ Tarucha, {\em 
Excitation Spectra of Circular Few-Electron Quantum Dots}, Science 
{\bf 278}
1788--1792 
(1997)

\bibitem[14]{Me93} U.\ Merkt,  {\em Far-infrared spectroscopy of quantum 
dots}, 
Physica B {\bf 189}, 165--175 (1993)

\bibitem[15]{Heit97} D.\ Heitmann, K.\ Bollweg, V.\ Gudmundsson, T.\ 
Kurth, S.\ P.\ Riege, {\em Far-infrared spectroscopy of quantum vires 
and dots, breaking Kohn's theorem}, Physica E {\bf 1}, 204--210 
(1997) 

\bibitem[16]{Zhin97} N.\ B.\ Zhintev, R.\ C.\ Ashoori, L.\ N.\ Pfeiffer and 
K.\ 
W.\ West, {\em Periodic and Aperiodic Bunching in the Addition Spectra of 
Quantum Dots}, Phys.\ Rev.\ Lett.\  {\bf 79}, 2308--2311 (1997)


\bibitem[17]{F28}  V.\ Fock, {\em Bemerkung zur Quantelung des harmonischen
Oszillators in Magnetfeld}, 
Z. Phys. {\bf 47}, 446--448 (1928)

\bibitem[18]{D30}
C.G.\ 
Darwin, {\em The Diamagnetism of the Free Electron}, 
Proc.\ Cambr.\ Philos.\ Soc.\ {\bf 27}, 86--90 (1930)

\bibitem[19]{Kohn61} W.\ Kohn, {Cyclotron Resonance and the de 
Haas-van Alphen Oscillations of an Interacting Electron Gas}, 
Phys.\ Rev.\ {\bf 123}, 1242--1244 (1961)

\bibitem[20]{GoCho90} A.\ O.\ Govorov and A.\ V.\ Chaplik, 
{\em Magnetoabsorption at quantum points}, JETP Lett.\ {\bf 52}, 31-33 
(1990)

\bibitem[21]{MacDo93} A.\ H.\ MacDonald, S.\ R.\ Yang, M.\ D.\ Johnson,
{\em Quantum dots in strong magnetic fields: Stability criteria for the 
maximum density droplet}, Australian J.\ Phys.\ {\bf 46}, 345 (1993)


\bibitem[22]{MueKoon96}S.\ E.\ Koonin and H.\ M.\ Mueller, {\em 
Phase-Transitions in Quantum Dots}, Phys.\ Rev.\ B {\bf 54}, 14532--14539 
(1996)

\bibitem[23]{Fer97}M.\ Ferconi and G.\ Vignale, {\em Density functional 
theory 
of the phase diagram of maximum density droplets in two dimensional quantum 
dots in a magnetic field}, Phys.\ Rev.\ B {\bf 56}, 12108--12111  (1997)


\bibitem[24]{Ost98} T.\ H.\ Osterkamp, J.\ W.\ Janssen, L.\ P.\ 
Kouwenhouven, 
D.\ G.\ Austing, T.\ Honda and S.\ Tarucha, {\em Stability of the maximum 
density drop in quantum dots at high magnetic fields}, Preprint, http:// 
vortex. tn.tudelft.nl/mensen/leok/papers/ (1998)


\bibitem[25]{Tau95} M.\ Taut, {\em Two electrons in a homogeneous magnetic 
field: Particular analytical solutions}, J.\ Phys.\ A {\bf 27}, 
1045--1055; Corrigendum {\bf 27}, 4723--4724  (1994)


\bibitem[26]{QuirJoh93} L.\ Quiroga, D.\ R.\ Ardila and N.\ F.\ Johnson, 
{\em 
Spatial Correlation of Quantum Dot Electrons in a Magnetic Field}, Solid 
State 
Comm.\ {\bf 86}, 775--780 (1993)

\bibitem[27]{JohPay91} N.\ F.\ Johnson and M.\ C.\ Payne, {\em Exactly 
Solvable Models of Interacting Particles in a Quantum Dot}, Phys.\ Rev.\ 
Lett.\ {\bf 67}, 1157--1160 (1991)

\bibitem[28]{Shi91}V.\ Shikin, S.\ Nazin, D.\ Heitmann and T.\ Demel, {\em 
Dynamical response of quantum dots}, Phys.\ Rev.\ B {\bf 43}, 11903--11907 
(1991)

\bibitem[29]{EFK93} P.L.\ McEuen, N.S.\ Wingreen, E.B.\ Foxman, J.\ Kinaret, 
U.\ Meirav, M.A.\ Kastner,  and Y.\ Meir, {\em Coulomb interactions 
and the energy-level spectrum of a small electron gas}, Physica  B 
{\bf 189}, 70--79 (1993) 

\bibitem[30]{LL}  E.H.\ Lieb and M.\ Loss, unpublished section of a book on
stability of matter.

\bibitem[31]{VRK94} N.C.\ van der Vaart, M.P.\ de Ruyter van Steveninck, 
L.P.\ Kouwenhoven, A.T.\ Johnson, Y.V.\ Nazarov, and C.J.P.M.\ Harmans, 
{\em
Time-Resolved Tunneling of Single Electrons between Landau Levels in a
Quantum Dot}, Phys.\ Rev.\ Lett.\ {\bf 73}, 320--323 (1994)



\bibitem[32]{LSY94a} E.H.\ Lieb, J.P.\ Solovej and J.\ Yngvason, {\em 
Asymptotics of
Heavy Atoms in High Magnetic Fields: I. Lowest Landau Band Regions},
Commun.\ Pure Appl.\ Math.\ {\bf 47}, 513--591 (1994)

\bibitem[33]{LSY94b} E.H.\ Lieb, J.P.\ Solovej and J.\ Yngvason, {\em 
Asymptotics of
Heavy Atoms in High Magnetic Fields: II. Semiclassical Regions}, Commun.\
Math.\ Phys {\bf 161}, 77-124 (1994)

\bibitem[34]{Lvar}E.H. Lieb, {\em A Variational Principle for Many-Fermion
Systems}, Phys. Rev. Lett. {\bf 46}, 457--459; Erratum {\bf 47}, 69
(1981)


\bibitem[35]{LY88} E.H.\ Lieb and H.-T.\ Yau, {\em The stability and 
instability of
relativistic matter}, Commun.\ Math.\ Phys.\ {\bf 118}, 177--213 (1988)

 
 \bibitem[36]{ESa} L.\ Erd\H os and J.P.\ Solovej, {\em Semiclassical 
eigenvalue estimates for the Pauli operator with strong non-homogeneous 
magnetic fields. I. Non-asymptotic Lieb Thirring estimates}, Duke Math. J., 
to 
appear
 
 \bibitem[37]{ESb} L.\ Erd\H os and J.P.\ Solovej, {\em Semiclassical 
eigenvalue estimates for the Pauli operator with strong non-homogeneous 
magnetic fields. II. Leading order asymptotic estimates}, 
 Comm.\ Math.\ Phys.\ {\bf 188}, 599--656 (1997)



\bibitem[38]{L79} E.\ H.\ Lieb, {\em A lower bound for Coulomb energies}, 
Phys.\ Lett.\ {\bf 70} A, 444--446 (1979)

\bibitem[39]{Ba92} V.\ Bach, {\em Error Bound for the Hartree-Fock Energy of 
Atoms and Molecules}, Commun.\ Math.\ Phys.\ {\bf 147}, 527--548 (1992)


\bibitem[40]{Yang93} S.\ R.\ Eric Yang, A.\ H.\ MacDonald and M.\ D.\ 
Johnson, 
{\em Addition Spectra of Quantum Dots in Strong Magnetic Fields}, Phys.\ 
Rev.\ 
Lett.\ {\bf 71}, 3194--3197 (1993)




\bibitem[41]{Wagn92} M.\ Wagner, U.\ Merkt and A.\ V.\ Chaplik, {\em 
Spin-singlet-spin-triplet oscillations in quantum dots}, Phys.\ Rev.\ B {\bf 
45}, 1951--1954 (1992)



\bibitem[42]{Pfann94}D.\ Pfannkuche, V.\ Gudmundsson and P.\ A.\ Maksym, 
{\em 
Comparison of a Hartree, a Hartree-Fock, and an exact treatment of quantum 
dot 
helium}, Phys.\ Rev.\ B {\bf 47}, 2244--2250 (1993)

\bibitem[43]{Pfann95} D.\ Pfannkuche and S.\ E.\ Ulloa, {\em Selection Rules 
for 
Transport Spectroscopy of Few-Electron Quantum Dots}, Phys.\ Rev.\ Lett.\ 
{\bf 
74}, 1194--1197 (1995)

\bibitem[44]{Meza97}L.\ Meza-Montes, S.\ E.\ Ulloa and D.\ Pfannkuche, {\em 
Electron interactions, classical instability, and level statistics in 
quantum 
dots}, Physica E {\bf 1}, 274--280 (1997)


\bibitem[45]{Eto97} M.\ Eto, {\em Electronic Structures of Few Electrons 
in a Quantum Dot under Magnetic Fields}, Jpn.\ J.\ Appl.\ Phys.\ {\bf 
36}, 3924--3927 (1997)


\bibitem[46]{Anis98} E.\ Anisimovas and A.\ Matulis, {\em Energy Spectra of 
Few-Electron Quantum Dots}, 
Jour. Phys. Cond. Mat.  {\bf 10}, 601--615 (1998)

\bibitem[47]{DiNaz97} M.\ Dineykhan and R.\ G.\ Nazmitdinov, {\em 
Two-Electron Quantum Dot in Magnetic Field: Analytical Results},
Phys.\ Rev.\ B {\bf 55}, 13707-13714 (1997)

\bibitem[48]{HarNie98a}A.\ Harju, V.\ A.\ Sverdlov and R.\ M.\ Nieminen, 
{\em 
Variational wave function for a quantum dot in a magnetic field: A quantum 
Monte-Carlo study}, Europhys.\ Lett.\, {\bf 41}, 407-412 (1998)

\bibitem[49]{HarNie98b}A.\ Harju, V.\ A.\ Sverdlov and R.\ M.\ Nieminen, 
{\em
Many-Body Wave Function for a Quantum Dot in a Weak magnetic Field}, 
Preprint, Univ. of Helsinki (1998)



\bibitem[50]{Fer94}M.\ Ferconi and G.\ Vignale. {\em Current density 
functional 
theory of quantum dots in a magnetic field}, Physical Review B {\bf 50}, 
14722--14725 (1994)


\bibitem[51]{Pi98}M.\ Pi, M.\ Barranco, A.\ Emperador, E.\ Lipparini and 
Ll.\ 
Serra, {\em Current Density Functional approach to large quantum dots in 
intense magnetic fields}, Phys.\ rev.\ B {\bf 57}, 14783--14792 (1998)

\bibitem[52]{Hein98} O.\ Heinonen, J.\ M.\ Kinaret and M.\ D.\ Johnson, {\em 
Ensemble Density Functional Approach to Charge-Spin Textures in 
Inhomogeneous 
Quantum Systems}, Phys.\ Rev.\ B (1998)










\end{thebibliography}
\end{document}